\documentclass[12pt]{article}
\usepackage[latin1]{inputenc}
\usepackage{amssymb}
\usepackage{hyperref}
\usepackage{amsmath}
\usepackage{latexsym}
\usepackage{times}
\usepackage{cite}
\usepackage{amssymb}
\usepackage{amsfonts}
\usepackage{amscd}
\usepackage{xcolor}
\usepackage{amstext,amsmath,amssymb,amsfonts}
\newtheorem{theorem}{Theorem}

\newtheorem{definition}[theorem]{Definition}

\newtheorem{lemma}[theorem]{Lemma}

\newtheorem{proposition}[theorem]{Proposition}
\newtheorem{remark}[theorem]{Remark}

\newcommand{\beq}{\begin{eqnarray}}
\newcommand{\eeq}{\end{eqnarray}}
\newcommand{\beqs}{\begin{eqnarray*}}
	\newcommand{\eeqs}{\end{eqnarray*}}
\newcommand{\bpro}{\begin{pro}}
	\newcommand{\epro}{\end{pro}}
\newcommand{\blem}{\begin{lem}}
	\newcommand{\elem}{\end{lem}}
\newcommand{\bdfn}{\begin{dfn}}
	\newcommand{\edfn}{\end{dfn}}
\newcommand{\bcor}{\begin{cor}}
	\newcommand{\ecor}{\end{cor}}
\newcommand{\bthm}{\begin{thm}}
	\newcommand{\ethm}{\end{thm}}
\newcommand{\bex}{\begin{ex}}
	\newcommand{\eex}{\end{ex}}
\newcommand{\brmk}{\begin{rmk}}
	\newcommand{\ermk}{\end{rmk}}
\newcommand{\bpr}{\begin{pr}}
	\newcommand{\epr}{\end{pr}}
\newcommand{\benum}{\begin{enumerate}}
	\newcommand{\eenum}{\end{enumerate}}
\newcommand{\bitem}{\begin{itemize}}
	\newcommand{\eitem}{\end{itemize}}

\newcommand{\cqfd}{\hfill{\square}}
\chardef\bslash=`\\
\numberwithin{equation}{section}
\numberwithin{table}{section}
\numberwithin{theorem}{section}

\begin{document}
	\begin{flushright}
		ICMPA-MPA/2018
	\end{flushright}
	\begin{center}
		{\Large { $\mathcal{R}(p,q)-$deformed conformal Virasoro
				 algebra} }\\
		\vspace{0,5cm}
		Mahouton Norbert Hounkonnou$^{*}$ and Fridolin Melong\\
		\vspace{0.5cm}
		{\em International Chair in Mathematical Physics
			and Applications}
		{\em (ICMPA-UNESCO Chair), }
		{\em University of Abomey-Calavi,}
		{\em 072 B.P. 50 Cotonou, Republic of Benin}\\	
{\em E-mails: norbert.hounkonnou@cipma.uac.bj, (with copy to hounkonnou@yahoo.fr);}
{\em fridomelong@gmail.com}
	\end{center}
	\today
	
	\vspace{0.5 cm}
	{\it Abstract}

This paper addresses an $\mathcal{R}(p,q)$-deformed conformal Virasoro algebra with an arbitrary conformal dimension $\Delta.$ 
Wellknown  deformations constructed in the literature
 are deduced as particular cases.
  Then,  the special case of
 the conformal dimension $\Delta=1$ is elucidated for its interesting properties. 
The $\mathcal{R}(p,q)-$KdV equation, associated with the deformed Virasoro algebra, is  also derived and discussed.
Finally, the $(p,q)$-deformed energy-momentum tensor,
 consistent with the  central extension term,  
 is computed and  analyzed.

	{\noindent
		{\bf Keywords.}
		$\mathcal{R}(p,q)-$ calculus; Virasoro algebra;  central extension; conformal dimension; KdV equation; energy-momentum tensor.	}
	\tableofcontents

\section{Introduction}
Quantum groups and algebras, in both their construction and representation theory, with or without  deformation, are exciting areas of mathematics, originated from mathematical physics of field theory and statistical mechanics.   Connected with  many  other parts of mathematics, these topics were broadly developed over the last decades,  and remain   today an area of prolific
research activities. This paper deals with a construction of the deformed Virasoro algebra on the line of recently introduced ${\mathcal R}(p,q)-$deformed quantum algebras \cite{HB1}.
 To cite a few relevant works related to the deformation of Virasoro algebra, see 
 \cite{CILPP},\cite{CKL} (and references therein).  The Virasoro algebra with a conformal dimension $\Delta$ is related to the Korteweg-de-Vries (KdV) integrable systems, and  plays an important role in physics. This motivated the series of works devoted these last years  to its  deformation  and generalization \cite{CKL},\cite{CPP},\cite{Hounkonnou:2015laa}. 
In the same vein, were also realized  
$q$-deformed Virasoro algebras with  conformal parameter $\Delta$,  multiplicative and comultiplication rule for  deformed generators \cite{DS},  $q$-deformed central extension term \cite{AS},\cite{CILPP},\cite{CEP}, and related $q$-deformed KdV equation \cite{CPP}.
The relevance of $\Delta =0, 1/2, 1,$  and  $q$-deformed energy-momentum tensor associated with the $q$-deformed central extension term\cite{CILPP} was  highlighted. The Virasoro algebras with a conformal dimension  $\Delta$ and two deformation parameters, (also called conformal $(p,q)$-Virasoro algebras), were  also investigated by some authors, taking into account    properties of  comultiplication,   deformed nonlinear  $(p,q)$-KdV-equation 
and  deformed central extension term \cite{CJ}.
Recently, the  generalized  Virasoro algebra, and related algebraic and  hydrodynamics properties were studied in \cite{Hounkonnou:2015laa}.

Motivated by  the generalization of the well- known $(p,q)$-deformation in 
\cite{HB}, we address in this paper  the $\mathcal{R}(p,q)$-deformed conformal Virasoro algebra with an arbitrary dimension $\Delta,$ with a particular emphasis on properties induced  by $\Delta=1$ and  $2.$

This paper is organized as follows. Main definitions and notations 
are briefly  recalled in Section $2.$  In  Section $3,$ the  $\mathcal{R}(p,q)$-deformation of the conformal Virasoro algebra with an arbitrary conformal dimension $\Delta$ is performed. The $\mathcal{R}(p,q)$-deformed Jacobi identity is  discussed, the deformation of the central extension term depending on the meromorphic function $\mathcal{R}$ is obtained,  and the $\mathcal{R}(p,q)$-deformed conformal Virasoro algebra is constructed. Interesting results are derived for deduced known  particular deformations \cite{Chakrabarti&Jagan},\cite{HNN},\cite{JS}.  Section $4$ describes the case of
 the dimension $\Delta=1$. 
Then,  an $\mathcal{R}(p,q)$-nonlinear differential equation
 is obtained and  linked to  the $\mathcal{R}(p,q)$-deformed conformal  Virasoro algebra. In Section $5$, we  compute  the $(p,q)$-deformed energy-momentum tensor, which is
 consistent with the $(p,q)$-deformed central extension term.
 Section $6$ is devoted to concluding remarks.  
\section{Preliminaries: basic definitions and notations}
In this section, we briefly recall main definitions, notations and known results used in the sequel.

Let us consider an arbitrary conformal dimension $\Delta$ \cite{CJ}. A field $\phi_{\Delta}(z)$ with the conformal dimension $\Delta$ transforms under an infinitesimal coordonate transformation \cite{CILPP} $z\mapsto z + f(z)$ as \begin{equation}
\delta_{f(z)}\phi_{\Delta}(z) = f(z)^{1-\Delta}\partial_z(f(z)^{\Delta}\phi_{\Delta}(z)).
\end{equation}

For \begin{equation}\label{ba}
f(z)=z^{1+n},
\end{equation}

 we obtain
\begin{equation}\label{a}
\delta_n\phi_{\Delta}(z) := g_n(z) = (z\partial_z + \Delta(n+1)-n)z^n\phi_{\Delta}(z),
\end{equation}
where the  generators $g_n$ satisfy the centerless Virasoro algebra, known under the name of  Witt algebra, $\mathcal W,$

\begin{equation}\label{b}
\left[g_n, g_m \right]=(m-n)g_{n+m}.
\end{equation}
 The centrally extended Virasoro algebra $\mathcal V$ is spanned by the generators obeying 
 \begin{equation}
 \left[g_n, g_m \right]=(m-n)g_{n+m} + \frac{c}{12}(n^3-n)\delta_{n+m,o},
 \end{equation}
with the property, for all $g_n \in \mathcal V,$
\begin{equation}
[g_n,c]=0.
\end{equation}
Let  $\mathcal{R}$ be a meromorphic function defined on $\mathbb{C}\times\mathbb{C}$ by \begin{equation}\label{r10}
	\mathcal{R}(u,v)= \sum_{s,t=-l}^{\infty}r_{st}u^sv^t,
	\end{equation}
converging in the complex disc $\mathbb{D}_{R}=\left\lbrace z\in\mathbb{C}/ |z|<R\right\rbrace,$
	where $r_{st}$ are complex numbers, $l\in\mathbb{N}\backslash\left\lbrace 0\right\rbrace$ and $R$ is the radius of convergence of the series (\ref{r10}).
	Let us consider the set of holomorphic functions $\mathcal{O}(\mathbb{D}_{R})$ defined on $\mathbb{D}_{R}$.

	\begin{definition}\cite{HB}
		Let $P$ and $Q$ be two linear operators on $\mathcal{O}(\mathbb{D}_{R})$.Then, for $\varPsi\in\mathcal{O}(\mathbb{D}_{R})$, we have \begin{equation}\label{r2}
		Q:\varPsi\longmapsto Q\varPsi(z)= \varPsi(qz),
		\end{equation}
		\begin{equation}\label{r3}
		P:\varPsi\longmapsto P\varPsi(z)= \varPsi(pz).
		\end{equation}
	\end{definition} 
\begin{definition}\cite{Chakrabarti&Jagan}
	The $(p,q)-$derivative and the $(p,q)-$number are defined, respectively, by: \begin{equation}\label{r4}
	D_{p,q} :\varPsi\longmapsto D_{p,q}\varPsi(z)=\frac{\varPsi(pz)-\varPsi(qz)}{z(p-q)},
	\end{equation}
	\begin{equation}
	[n]_{p,q}:=\frac{p^n-q^n}{p-q},
	\end{equation}
	where $p$ and $q,$ satisfying $0<q<p\leq 1,$ are two real numbers.
\end{definition}
\begin{definition}\cite{HB}
	The $\mathcal{R}(p,q)-$derivative is given by: 
	\begin{equation}\label{r5}
	D_{\mathcal{R}( p,q)}:=D_{p,q}\frac{p-q}{P-Q}\mathcal{R}( P,Q)=\frac{p-q}{p^{P}-q^{Q}}\mathcal{R}(p^{P},q^{Q})D_{p,q},
	\end{equation}
	where  $P,$ $Q$ are defined on $\mathcal{O}(\mathbb{D}_{R}),$ and $p,$  $q$ verifying $0<q<p\leq 1,$ are two real numbers. 

\end{definition}
\begin{definition}\cite{HB1}
	The  $\mathcal{R}(p,q)-$number and the  $\mathcal{R}(p,q)-$ factorials are defined, respectively, as follows:
	\begin{enumerate}
		\item \begin{equation}
		[n]_{\mathcal{R}(p,q)}:=\mathcal{R}(p^n,q^n)\quad\mbox{for}\quad n\geq 0,
		\end{equation}
		\begin{equation}\label{s0}
		[n]!_{\mathcal{R}(p,q)}:=\left \{
		\begin{array}{l}
		1\quad\mbox{for}\quad n=0\\
		\\
		\mathcal{R}(p,q)\cdots\mathcal{R}(p^n,q^n)\quad\mbox{for}\quad n\geq 1,
		\end{array}
		\right .
		\end{equation}
		where $p$ and $q,$ satisfying $0<q<p\leq 1,$ are two real numbers.
	\end{enumerate}
\end{definition}
\begin{lemma}\label{L1}\cite{HB1}
 	For $\varphi(z)=z^n \in \mathcal{O}(\mathbb{D}_{R})$, the $\mathcal{R}(p,q)$- derivative is given by:
 	
 	\begin{eqnarray}\label{r7}
 	D_{\mathcal{R}( p,q)}z^n&=& z^{-1}[z\partial_z]_{p,q}z^n\frac{p-q}{p^{-n}-q^{-n}}\mathcal{R}( p^n,q^n).
 	\end{eqnarray}
\end{lemma}

\begin{proposition}\cite{HB}\label{p1}
	Let $P$, $Q$ be two linear operators and $p$, $q$ two real numbers, then we have the following relations
	
	\begin{equation}
	P=p^{z\partial_z}\quad\mbox{and}\quad Q=q^{z\partial_z}.
	\end{equation}
\end{proposition}
\begin{definition}\cite{HB}
 The Jaganathan-Srinivasa $(p,q)$ factors and  $(p,q)$-factorials are given  by 
\begin{equation*}
[x]_{p,q}=(p-q)^{-1}(p^x-q^x),
\end{equation*}
and
\begin{equation*}
[x]!_{p,q}=\left \{
\begin{array}{l}
\frac{\left((p,q);(p,q)\right)_x }{(p-q)^x} \quad\mbox{for} \quad x\geq 1\\
\\
1 \quad\mbox{for} \quad x=0,
\end{array}
\right .
\end{equation*}
respectively,
 where  $p,$ $q$ satisfying $0<q<p\leq 1$  
are two real numbers .
\end{definition}
The  $(p,q)-$deformation of quantum algebras   gives rise to a realization of a $(p,q)$-oscillator \cite{Chakrabarti&Jagan}. The main properties of this deformation can be found in
\cite{Chakrabarti&Jagan} and \cite{HB}.
\begin{definition}
 The \textbf{Chakrabarty and Jagannathan} $(p^{-1},q)$ factors and  $(p^{-1},q)$-factorials are given  by :
\begin{equation*}
[x]_{p^{-1},q}=(p^{-1}-q)^{-1}(p^{-x}-q^x),
\end{equation*}
and
\begin{equation*}
[x]!_{p^{-1},q}=\left \{
\begin{array}{l}
1 \quad\mbox{for} \quad x=0\\
\\
\frac{\left((p^{-1},q);(p^{-1},q)\right)_x }{(p^{-1}-q)^x} \quad\mbox{for} \quad x\geq 1,
\end{array}
\right .
\end{equation*}
where  $p,$ $q$ satisfying $0<q<p\leq 1$  
are two real numbers .
\end{definition}

\begin{definition}\cite{HB}
 The \textbf{ (p,q)-Quesne} factors and factorials are given  by :
	\begin{equation*}
	[n]^Q_{p,q}=(-p^{-1}+q)^{-1}(p^n-q^{-n}),
	\end{equation*}
	and
	\begin{equation*}
	[n]^Q!_{p,q}=\left \{
	\begin{array}{l}
	1 \quad\mbox{for} \quad n=0\\
	\\
	\frac{\left((p,q^{-1});(p,q^{-1})\right)_n }{(-p^{-1}+q)^n} \quad\mbox{for} \quad n\geq 1.
	\end{array}
	\right .
	\end{equation*}
\end{definition}
The above mentioned deformations are recoverable in the work  \cite{Hounkonnou&Ngompe07a} as particular cases. Finally, let $p,q,\nu,\mu$ be four real numbers such that $0<pq<1$ , $p^{\mu}< q^{\nu -1}$, $p>1,$ and $g$  a  real function of two-parameter deformation $p$ and $q$ verifying $g(p,q)\longrightarrow 1$ as $(p,q)\longrightarrow (1,1)$. 
 The $(p,q,\mu,\nu,g)$- factorial and  
the $(p,q,\mu,\nu,g)$-number are  given by \cite{Hounkonnou&Ngompe07a}:
\begin{equation}\label{s4}
[n]^{\mu,\nu}_{p,q,g}!\equiv\left \{
\begin{array}{l}
[n]^{\mu,\nu}_{p,q,g}[n-1]^{\mu,\nu}_{p,q,g}\cdots [2]^{\mu,\nu}_{p,q,g}[1]^{\mu,\nu}_{p,q,g} \quad\mbox{if} \quad n=1,2,\cdots\\
\\
1 \quad\mbox{if} \quad n=0,
\end{array}
\right .
\end{equation} 
and 
\begin{equation}\label{H1}
[n]^{\mu,\nu}_{p,q,g}=g(p,q)\frac{q^{n\nu}}{p^{n\mu}}\frac{p^n-q^{-n}}{q-p^{-1}},
\end{equation}
respectively.
\section{$\mathcal{R}(p,q)$-deformed conformal algebra}
In this section, we derive  an $\mathcal{R}(p,q)-$extension of a conformal algebra with an arbitrary conformal dimension $\Delta$ \cite{CJ}. The $\mathcal{R}(p,q)$-generators are computed using the analogue of the $\mathcal{R}(p,q)-$Leibniz rule. The
 $\mathcal{R}(p,q)$-deformed Witt  and Virasoro algebras are built and discussed. A  result deduced by  Chaichian {\it et al} in \cite{CILPP} is recovered in a specific case.  

\subsection{General construction}
Let us consider an infinitesimal $\mathcal{R}(p,q)$-transformation for a field $\phi_{\Delta}(z)$ as: 

 \begin{equation}\label{c}
\delta^{\mathcal{R}(p,q)}_{f(z)}\phi_{\Delta}(z) := f(z)^{1-\Delta}D_{\mathcal{R}(p,q)}(f(z)^{\Delta}\phi_{\Delta}(z)).
\end{equation}
 
For $f(z):=z^{1+n}$, we define the $\mathcal{R}(p,q)$-Virasoro generators $\mathcal{L}^{\Delta}_n$ by:

 \begin{equation}\label{e1}
\mathcal{L}^{(\Delta)}_n\phi_{\Delta}(z):=z^{(1+n)(1-\Delta)}D_{\mathcal{R}(p,q)}(z^{\Delta(1+n)}\phi_{\Delta}(z)).
\end{equation}

According to the $\mathcal{R}(p,q)$-derivative (\ref{r5}), we use the following lemma.

\begin{lemma}\label{Lr}
	 The $\mathcal{R}(p,q)$-Leibniz rule is given by: 
	
	\begin{equation}
	D_{\mathcal{R}( p,q)}(\varepsilon(z)^{\Delta}\phi_{\Delta}(z))=Kh(P,Q)z^{-1}\left\lbrace \varepsilon(pz)^{\Delta}\phi_{\Delta}(pz)-\varepsilon(qz)^{\Delta}\phi_{\Delta}(qz)\right\rbrace , 
	\end{equation}
	where $K=(p-q)^{-1}$ ,

	\begin{equation}
	h(P,Q)=\frac{p-q}{p^{P}-q^{Q}}\mathcal{R}(p^{P},q^{Q}).
	\end{equation} 
	 $\phi_{\Delta}$ is an arbitrary primary  field with the conformal dimension $\Delta$, $\varepsilon \in\mathcal{O}(D_R).$ $P,$ $Q$ are two linear operators on $\mathcal{O}(D_R),$ while   $p,$ $q$   are two real numbers satisfying $0<q<p\leq 1.$
\end{lemma}
{\it Proof:} Let us consider two functions $f(z)$ and $g(z)$ belonging to $\mathcal{O}(D_R).$ Then using the $\mathcal{R}(p,q)$-derivative, we get

\begin{equation}
D_{\mathcal{R}( p,q)}(f(z)g(z))=\frac{p-q}{p^{P}-q^{Q}}\mathcal{R}(p^{P},q^{Q})D_{p,q}((f(z)g(z)).
\end{equation}
Setting $f(z)=\varepsilon(z)^{\Delta}$,  $g(z)=\phi_{\Delta}(z)$ and using the $(p,q)$-Leibniz rule \cite{HB},  the result is immediately obtained. $\cqfd$ 

\begin{proposition}\label{l1} 
The $\mathcal{R}(p,q)$-generators of the $\mathcal{R}(p,q)$-deformed Virasoro algebra are given by:

	\begin{eqnarray}\label{P2}
\mathcal{L}^{(\Delta)}_n\phi_{\Delta}(z)&=& \left[ z\partial_z + \Delta(n+1) -n \right]z^{n}\omega^{\Delta}_n(p,q)\phi_{\Delta}(z),
	\end{eqnarray}	
	where
	\begin{equation}\label{pa}
	w^{\Delta}_n(p,q) = (p-q)(p^{-\Delta(n+1)}-q^{-\Delta(n+1)})^{-1}\mathcal{R}(p^{\Delta(n+1)},q^{\Delta(n+1)}).
	\end{equation}
\end{proposition}
{\it Proof:}
The proof uses (\ref{e1}), and  Lemmas \ref{L1} and  \ref{Lr}.
$\cqfd$
\begin{remark}
	\item (i) For the choice of the meromorphic function $\mathcal{R}(x,y)=(p-q)^{-1}(\frac{1}{x}-\frac{1}{y})$, we get the $(p,q)$-Virasoro generators given by Chakrabarti et {\it al}\cite{CJ}.
	\item(ii) The $\mathcal{R}(p,q)$-Virasoro generators $\mathcal{L}^{(\Delta)}_n$ can be expressed in  terms of the $(p,q)$-Virasoro generators $e^{\Delta
		}_n,$ i.e. \begin{equation}\label{r1}
		\mathcal{L}^{(\Delta)}_n = w^{(\Delta)}_n(p,q)e^{\Delta
		}_n
		\end{equation}
\end{remark}
\begin{proposition}\label{P3}
 The $\mathcal{R}(p,q)$-deformed generators $\mathcal{L}^{(\Delta)}_n$ satisfy the following commutation relation:
\begin{eqnarray}\label{P1}
[\mathcal{L}^{(\Delta)}_n , \mathcal{L}^{(\Delta)}_m]_{\tilde{X}_{\Delta},\tilde{Y}_{\Delta}}&=& \tilde{X}_{\Delta}\mathcal{L}^{(\Delta)}_n\mathcal{L}^{(\Delta)}_m -\tilde{Y}_{\Delta}\mathcal{L}^{(\Delta)}_m \mathcal{L}^{(\Delta)}_n\nonumber\\
&=& K\left\lbrace p^{N_{\Delta}}(\tilde{X}_{\Delta}p^{-n}-\tilde{Y}_{\Delta}p^{-m}) - q^{N_{\Delta}}(\tilde{X}_{\Delta}q^{-n}-\tilde{Y}_{\Delta}q^{-m}) \right\rbrace\nonumber\\
&\times& \mathcal{L}^{(\Delta)}_{n+m},
\end{eqnarray} 

where 
\begin{equation}\label {3}
\left \{
\begin{array}{l}
{\tilde{X}_{\Delta}}= (pq)^{n}\frac{[n(\Delta -1)][\Delta m]}{[n][m]}\frac{w^{\Delta}_{n+m}(p,q)}{w^{\Delta}_n(p,q)w^{\Delta}_m(p,q)} \\
\\
{\tilde{Y}_{\Delta}}=(pq)^{m}\frac{[m(\Delta -1)][\Delta n]}{[n][m]}\frac{w^{\Delta}_{n+m}(p,q)}{w^{\Delta}_n(p,q)w^{\Delta}_m(p,q)}\\
\\
N_{\Delta} =z\partial_z + \Delta\\
\\
K=(p-q)^{-1}.
\end{array}
\right. 
\end{equation} 
\end{proposition}
{\it Proof:} It is obtained by a straightforward computation.
$\cqfd$

It is worth mentioning the following  relevant particular deformed quantum algebras together with their  conformal characterisation and properties, which are derived  from the above developed formalism:
\begin{enumerate}
{	\item  For
 $\mathcal{R}(x,y) = (p-q)^{-1}(\frac{1}{x}- \frac{1}{y}),$ and  the limit $(p,q \longrightarrow 1)$ in (\ref{P1}), we get  the results given by Chakrabarti et {\it al} \cite{CJ} and 
 Chaichian et {\it al} \cite{CILPP}, respectively. }  
	\item The \textbf{Jagannathan-Srinivasa} generators can be obtained by taking $\mathcal{R}(s,t)=(p-q)^{-1}(s-t)$ leading to generators $\mathcal{L}^{(\Delta)}_n$ acting on the conformal field $\phi_{\Delta}$ as:
	
	\begin{equation}\label{A1}
	\mathcal{L}^{(\Delta)}_n\phi_{\Delta}(z)=-(pq)^{\Delta(n+1)} [z\partial_z + \Delta(1+n) -n ]z^{n}\phi_{\Delta}(z),
	\end{equation}
	and satisfying the  algebraic structure (\ref{P1}) with
	\begin{equation}
	\tilde{X}_{\Delta}= -(pq)^{n-\Delta}\frac{[n(\Delta -1)][\Delta m]}{[n][m]}
	\end{equation}
	and \begin{equation}
	\tilde{Y}_{\Delta}= -(pq)^{m-\Delta}\frac{[m(\Delta -1)][\Delta n]}{[n][m]}.
	\end{equation}
	\item The deformed \textbf{Chakrabarti-Jagannathan algebra} \cite{Chakrabarti&Jagan} corresponds to the  choice $\mathcal{R}(s,t)=(p^{-1}-q)^{-1}s^{-1}(1-st)$ yielding the generators $\tilde{L}_n$ acting as:
	\begin{equation}
	\tilde{L}_n\phi_{\Delta}(z)=-(pq)^{\Delta(n+1)}\frac{[\Delta(n+1)]_{p^{-1},q}}{[\Delta(n+1)]_{p,q}}[z\partial_z + \Delta(1+n) -n ]z^{n}\phi_{\Delta}(z), 
	\end{equation} 
	and obeying  the  algebraic structure (\ref{P1}) with 
	\begin{eqnarray*}
	\tilde{X}_{\Delta}&=& -(pq)^{n-\Delta}\frac{[n(\Delta -1)][\Delta m]}{[n][m]}\frac{[\Delta(n+m+1)]_{p^{-1},q}}{[\Delta(n+m+1)]}\cr
&& \frac{[\Delta(n+1)][\Delta(m+1)]}{[\Delta(n+1)]_{p^{-1},q}[\Delta(m+1)]_{p^{-1},q}},
	\end{eqnarray*}
and 	\begin{eqnarray*}
	\tilde{Y}_{\Delta}&=& -(pq)^{m-\Delta}\frac{[m(\Delta -1)][\Delta n]}{[n][m]}\frac{[\Delta(n+m+1)]_{p^{-1},q}}{[\Delta(n+m+1)]}\cr &&
 \frac{[\Delta(n+1)][\Delta(m+1)]}{[\Delta(n+1)]_{p^{-1},q}[\Delta(m+1)]_{p^{-1},q}}.
\end{eqnarray*}
\item Putting $\mathcal{R}(s,t)=((-p^{-1}+q)t)^{-1}(st-1)$, we obtain the \textbf{generalized Quesne} deformed  algebra \cite{Hounkonnou&Ngompe07a} with the  generator $\tilde{L}_n$  acting as: 
\begin{equation}
\tilde{L}_n\phi_{\Delta}(z)=-(pq)^{\Delta(n+1)}\frac{[\Delta(n+1)]^{Q}_{p,q}}{[\Delta(n+1)]_{p,q}}[z\partial_z + \Delta(1+n) -n ]z^{n}\phi_{\Delta}(z),
\end{equation}
and  satisfying the relation (\ref{P1}) with 
\begin{eqnarray*}
	\tilde{X}_{\Delta}= -(pq)^{n-\Delta}\frac{[n(\Delta -1)][\Delta m]}{[n][m]}\frac{[\Delta(n+m+1)]^Q_{p,q}}{[\Delta(n+m+1)]}
 \frac{[\Delta(n+1)][\Delta(m+1)]}{[\Delta(n+1)]^Q_{p,q}[\Delta(m+1)]^Q_{p,q}},
\end{eqnarray*}
and 	\begin{eqnarray*}
	\tilde{Y}_{\Delta}= -(pq)^{m-\Delta}\frac{[m(\Delta -1)][\Delta n]}{[n][m]}\frac{[\Delta(n+m+1)]^Q_{p,q}}{[\Delta(n+m+1)]}
 \frac{[\Delta(n+1)][\Delta(m+1)]}{[\Delta(n+1)]^Q_{p,q}[\Delta(m+1)]^Q_{p,q}}.
\end{eqnarray*}
\item Taking $\mathcal{R}(x,y)=g(p,q)\frac{y^{\nu}}{x^{\mu}}\frac{xy -1}{(q-p^{-1})y}$, 
(see properties in  \cite{HB}), we obtain the deformed \textbf{Hounkonnou-Ngompe} generalized 
algebra induced by the
generators $\tilde{L}_n$ such that: 
\begin{equation}
\tilde{L}_n\phi_{\Delta}(z)=-(pq)^{\Delta(n+1)}\frac{[\Delta(n+1)]^{\mu,\nu}_{p, q, g}}{[\Delta(n+1)]_{p,q}}[z\partial_z + \Delta(1+n) -n ]z^{n}\phi_{\Delta}(z),
\end{equation}
and  the commutation relation (\ref{P1})  with 
\begin{eqnarray*}
	\tilde{X}_{\Delta}&=& -(pq)^{n-\Delta}\frac{[n(\Delta -1)][\Delta m]}{[n][m]}\frac{[\Delta(n+m+1)]^{\mu,\nu}_{p,q,g}}{[\Delta(n+m+1)]}\cr
&& \frac{[\Delta(n+1)][\Delta(m+1)]}{[\Delta(n+1)]^{\mu,\nu}_{p,q,g}[\Delta(m+1)]^{\mu,\nu}_{p,q,g}},
\end{eqnarray*}
and 
	\begin{eqnarray*}
	\tilde{Y}_{\Delta}&=& -(pq)^{m-\Delta}\frac{[m(\Delta -1)][\Delta n]}{[n][m]}\frac{[\Delta(n+m+1)]^{\mu,\nu}_{p,q,g}}{[\Delta(n+m+1)]}\cr
&& \frac{[\Delta(n+1)][\Delta(m+1)]}{[\Delta(n+1)]^{\mu,\nu}_{p,q,g}[\Delta(m+1)]^{\mu,\nu}_{p,q,g}}.
\end{eqnarray*}
\item The generalized algebra derived by \textbf{Hounkonnou-Bukweli} in  \cite{HB1}
with
\begin{equation}\label{Jd}
[n]^{\mu,\nu}_{p,q,g} = g(p,q)\frac{q^{\nu n}}{p^{\mu n}}[n]^{Q}_{p,q,g},
\end{equation} 
 yields the deformed conformal generators $\tilde{L}_n$ 

\begin{equation}
\tilde{L}_n=-g(p,q)\left( (pq)\frac{q^{\nu}}{p^{\mu}}\right)^{\Delta(n+1)} \frac{[\Delta(n+1)]^Q_{p, q, g}}{[\Delta(n+1)]_{p,q}}[z\partial_z + \Delta(1+n) -n ]z^{n},
\end{equation}
which  satisfy  the relation (\ref{P1}) with
\begin{eqnarray}
	\tilde{X}_{\Delta}&=& -(pq)^{n-\Delta}g(p,q)^{-1}\left( \frac{q^{\nu }}{p^{\mu }}\right)^{n\Delta} \frac{[n(\Delta -1)][\Delta m]}{[n][m]}\nonumber\\
	&\times& \frac{[\Delta(n+m+1)]^Q_{p,q,g}}{[\Delta(n+m+1)]}\frac{[\Delta(n+1)][\Delta(m+1)]}{[\Delta(n+1)]^Q_{p,q,g}[\Delta(m+1)]^Q_{p,q,g}},
\end{eqnarray}
and 	\begin{eqnarray}
	\tilde{Y}_{\Delta}&=& -(pq)^{m-\Delta}g(p,q)^{-1}\left( \frac{q^{\nu }}{p^{\mu }}\right)^{n\Delta}\frac{[m(\Delta -1)][\Delta n]}{[n][m]}\nonumber\\
	&\times& \frac{[\Delta(n+m+1)]^Q_{p,q,g}}{[\Delta(n+m+1)]}\frac{[\Delta(n+1)][\Delta(m+1)]}{[\Delta(n+1)]^Q_{p,q,g}[\Delta(m+1)]^Q_{p,q,g}}.
\end{eqnarray}

\item The deformation of the \textbf{Chaichian et {\it al}} \cite{CILPP} Virasoro algebra is performed by considering two parameters, $p>0$ and $q>0$,
\begin{equation}\label{e9}
\lambda := \frac{\sqrt{pq}}{pq}\quad\mbox{and}\quad \theta:=\sqrt{\frac{p}{q}},
\end{equation}
with the numbers

\begin{equation}\label{e12}
[x]=\lambda^{1-x}[x]_{\theta},
\end{equation}
and
\begin{equation}
[x]_{\theta}= (\theta - \theta^{-1})^{-1}(\theta^{x} - \theta^{-x}).
\end{equation}

\end{enumerate}
	The associated $\mathcal{R}(p,q)$-deformed generators $\tilde{L}_n$  given by:
	\begin{equation}
	\tilde{L}_n\phi_{\Delta}(z)=T^{\Delta}_n(\lambda)\left[ z\partial_z + \Delta(1+n) -n\right] _{\theta}z^nw_n^{\Delta}(\theta,\theta^{-1})\phi_{\Delta}(z),
	\end{equation}
	where
	\begin{equation}
	T^{\Delta}_n(\lambda)=\lambda^{n-1-\Delta(2n+1)}\mathcal{R}(\lambda^{-\Delta(n+1)},\lambda^{-\Delta(n+1)}),
	\end{equation} 
satisfy the commutation relation:
		
	\begin{equation}\label{e13}
	[\tilde{L}_n , \tilde{L}_m]_{x,y} = K_\theta\left\lbrace \theta^{N_{\Delta}}(x\theta^{-n}-y\theta^{-m})-\theta^{-N_{\Delta}}(x\theta^{n}-y\theta^{m}) \right\rbrace \tilde{L}_{n+m},
	\end{equation} 
	where 
	
	\begin{equation}\label {e14}
	\left \{
	\begin{array}{l}
	\displaystyle
	x=\lambda^{(1-\Delta)(n+m)}\frac{[n(\Delta -1)]_{\theta}[\Delta m]_{\theta}}{[n]_{\theta}[m]_{\theta}}\frac{w^{\Delta}_{n+m}(\theta,\theta^{-1})}{w^{\Delta}_n(\theta,\theta^{-1})w^{\Delta}_m(\theta,\theta^{-1})}\\
	\\
	
	y=\lambda^{(1-\Delta)(n+m)}\frac{[m(\Delta -1)]_{\theta}[\Delta n]_{\theta}}{[n]_{\theta}[m]_{\theta}}\frac{w^{\Delta}_{n+m}(\theta,\theta^{-1})}{w^{\Delta}_n(\theta,\theta^{-1})w^{\Delta}_m(\theta,\theta^{-1})}\\
	\\
	K_\theta=(\theta-\theta^{-1})^{-1}.
	\end{array}
	\right. 
	\end{equation}
Let us now construct the $\mathcal{R}(p,q)-$ deformed conformal Witt algebra by  using the $\mathcal{R}(p,q)$-deformed conformal algebra (\ref{P1}). 
\begin{proposition}
	{Let $ \mathcal{L}^{(\Delta)}_n$ be the $\mathcal{R}(p,q)$-deformed generators, $m$ and $n$ two numbers belonging to the set of natural numbers. For  two real numbers $p,$ $q$ satisfying $0<q<p\leq 1,$   and all field $\phi_{\Delta}(z)$ of conformal dimension $\Delta$,  the following commutation  relation holds:}
\end{proposition}
\begin{equation}\label{e4}
[\mathcal{L}^{(\Delta)}_n, \mathcal{L}^{(\Delta)}_m]_{\hat{X},\hat{Y}}\phi_{\Delta}(z)=\mathcal{R}(p^{n-m},q^{n-m})\mathcal{L}^{(\Delta)}_{n+m}\phi_{\Delta}(z),
\end{equation}
with 

\begin{equation}\label {5}
\left \{
\begin{array}{l}
\displaystyle
\hat{X}=(p-q)\mathcal{R}(p^{n-m},q^{n-m}) \chi_{nm}(p,q)\\
\\

\hat{Y}=(p-q)\mathcal{R}(p^{n-m},q^{n-m}) \chi_{mn}(q,p).\\
\end{array}
\right. 
\end{equation}
and 
\begin{eqnarray}
\chi_{nm}(p,q)&=&\Big\lbrace p^{N_{\Delta}}(p^{-n}-\frac{[m(\Delta -1)][\Delta n]}{n(\Delta -1)][\Delta m]}\frac{q^{m-n}}{p^n}) - q^{N_{\Delta}}(q^{-n}\nonumber\\
&-&\frac{[m(\Delta -1)][\Delta n]}{n(\Delta -1)][\Delta m]}\frac{p^{m-n}}{q^n})\Big\rbrace^{-1},
\end{eqnarray}

{\it Proof:} It  uses the closed algebraic structure (\ref{P1}).  By setting

\begin{equation*}
\chi_{nm}(p,q)= \left\lbrace p^{N_{\Delta}}(p^{-n}-\frac{\tilde{Y}_{\Delta}}{\tilde{X}_{\Delta}}p^{-m}) - q^{N_{\Delta}}(q^{-n}-\frac{\tilde{Y}_{\Delta}}{\tilde{X}_{\Delta}}q^{-m}) \right\rbrace^{-1}.
\end{equation*}
we obtain:

\begin{eqnarray}\label{result0}
[\mathcal{L}^{(\Delta)}_n , \mathcal{L}^{(\Delta)}_m]_{\tilde{X}_{\Delta},\tilde{Y}_{\Delta}}&=& K\left\lbrace p^{N_{\Delta}}(\tilde{X}_{\Delta}p^{-n}-\tilde{Y}_{\Delta}p^{-m}) - q^{N_{\Delta}}(\tilde{X}_{\Delta}q^{-n}-\tilde{Y}_{\Delta}q^{-m}) \right\rbrace
\mathcal{L}^{(\Delta)}_{n+m}\nonumber\\
&=& K\tilde{X}_{\Delta}\left\lbrace p^{N_{\Delta}}(p^{-n}-\frac{\tilde{Y}_{\Delta}}{\tilde{X}_{\Delta}}p^{-m}) - q^{N_{\Delta}}(q^{-n}-\frac{\tilde{Y}_{\Delta}}{\tilde{X}_{\Delta}}q^{-m}) \right\rbrace
\mathcal{L}^{(\Delta)}_{n+m}.\nonumber\\
\end{eqnarray}

Replacing $\tilde{X}_{\Delta}$ and $\tilde{Y}_{\Delta}$ by their respective expressions yields:
\begin{eqnarray}
\chi_{nm}(p,q)&=&\Big\lbrace p^{N_{\Delta}}(p^{-n}-\frac{[m(\Delta -1)][\Delta n]}{n(\Delta -1)][\Delta m]}\frac{q^{m-n}}{p^n}) - q^{N_{\Delta}}(q^{-n}\nonumber\\
&-&\frac{[m(\Delta -1)][\Delta n]}{n(\Delta -1)][\Delta m]}\frac{p^{m-n}}{q^n})\Big\rbrace^{-1},
\end{eqnarray}
and
\begin{equation*}
\hat{X}=(p-q)\mathcal{R}(p^{n-m},q^{n-m})\chi_{nm}(p,q).
\end{equation*}
We explicitly compute $\hat{Y}$ in a similar way, and the required  result naturally comes by collecting different quantities in (\ref{result0}).
 $\cqfd$

\begin{remark}
	For a meromorphic function $R(x,y)=(p-q)^{-1}(x^{-1}-y^{-1})$ and the use of the transformation $q\longleftarrow p$, $p^{-1}\longleftarrow q$, the result given by Chakrabarti et {\it al}\cite{CJ} is recovered. 
\end{remark}

{The $\mathcal{R}(p,q)$-deformed Jacobi identity is derived in the following Lemma.
\begin{lemma}
	{Let $L_n$ be the $\mathcal{R}(p,q)$-deformed generators of the deformed conformal algebra, and $\hat{X}$, $\hat{Y}$ be the coefficients of the commutation relation (\ref{e4}). For all $n$, $m, $ and $k$ belonging to $\mathbb{N}$, 
	 the $\mathcal{R}(p,q)$-deformed Jacobi identity is given by:}
	 \begin{equation}\label{J3}
	 \displaystyle
	 \sum_{(u,v,l)\in\mathcal{C}(n,m,k)}\frac{(pq)^{-l}(p^u + q^u)}{\alpha^{\Delta}_{uvl}(p,q)}[L_u, [L_v , L_l]_{\hat{X},\hat{Y}}]_{\hat{X},\hat{Y}} =0,
	 \end{equation}
	  where $\mathcal{C}(n,m,k)$ denotes the
	  cyclic permutation of $(n,m,k)$. 
\end{lemma}
{\it Proof:} 
\begin{equation}\label{J14}
[L_n, [L_m , L_k]]=\alpha^{\Delta}_{nmk}[e_n, [e_m , e_k]],
\end{equation}
where
\begin{equation}\label{J15}
\alpha^{\Delta}_{nmk}(p,q)=\omega^{\Delta}_n(p,q)\omega^{\Delta}_m(p,q)\omega^{\Delta}_k(p,q),
\end{equation}
and  $e_n$ are the $(p,q)$-deformed generators.
By mimicking step by step
\cite{CJ}, we obtain:
\begin{equation}\label{J17}
(pq)^{-k}\frac{[2n]}{\alpha^{\Delta}_{nmk}[n]}[L_n, [L_m , L_k]_{\hat{X},\hat{Y}}]_{\hat{X},\hat{Y}}=(pq)^{-k}\frac{[2n]}{[n]} [e^{\Delta}_n, [e^{\Delta}_m , e^{\Delta}_k]_{R_{mk},S_{km}}]_{R_{n(m+k)},S_{(m+k)n}},
\end{equation}
and, finally, the result follows.
 $\cqfd$}

A central extension of the  $\mathcal{R}(p,q)$-deformed Witt algebra  (\ref{e4}) is governed by the commutation relations:

\begin{equation}\label{J18}
[\tilde{L}_n, \tilde{L}_m]_{\hat{X},\hat{Y}}=\mathcal{R}(p^{n-m},q^{n-m})\tilde{L}_{n+m} + \delta_{n+m,0}\tilde{C}^R_n(p,q),
\end{equation}
and

\begin{equation}\label{J19}
[\tilde{L}_k, \tilde{C}^R_n(p,q)]_{\hat{X}_k,\hat{Y}_k}=0,
\end{equation}
where 

\begin{equation}
\left \{
\begin{array}{l}
\displaystyle
\hat{X}_k=(p-q)\mathcal{R}(p^{k},q^{k}) \chi_{k0}(p,q)\\
\\

\hat{Y_k}=(p-q)\mathcal{R}(p^{k},q^{k}) \chi_{0k}(q,p),
\end{array}
\right. 
\end{equation}
with the central charge  given by the  factorization
\begin{equation}\label{eg}
\tilde{C}^R_n(p,q) = \tilde{\Gamma}(N_{\Delta})C^R_n(p,q),
\end{equation}
as shown in the sequel.
\begin{lemma}
 $\tilde{\Gamma}(N_{\Delta})$ satisfies the following identities:
	\begin{enumerate}
		\item For all $\Delta$ 
		\begin{equation}\label{g1}
		\tilde{X}_k\tilde{L}^{\Delta}_k\tilde{\Gamma}(N_{\Delta}) - \tilde{Y}_k\tilde{\Gamma}(N_{\Delta})\tilde{L}^{\Delta}_k=0.
		\end{equation}
		\item  For $\Delta=1/2$ 
		\begin{equation}\label{g2}
		(pq)^{k/2}\tilde{L}^{\Delta}_k\tilde{\Gamma}(N_{\Delta}) + \tilde{\Gamma}(N_{\Delta})\tilde{L}^{\Delta}_k=0.
		\end{equation}
		\item For $\Delta=2$
		\begin{equation}\label{g3}
		\tilde{L}^{\Delta}_k\tilde{\Gamma}(N_{\Delta}) - (p^{-k} + q^{-k})\tilde{\Gamma}(N_{\Delta})\tilde{L}^{\Delta}_k=0. 
		\end{equation}
	\end{enumerate}
\end{lemma}
{\it Proof:}
\begin{enumerate}
	\item From the relations (\ref{J19}) and (\ref{eg}), we obtain (\ref{g1}).
	 
	\item For $\Delta=1/2$, we obtain
	\begin{equation}
	\chi_{ko}(p,q)= \left\lbrace p^{N_{\Delta}}\left( p^{-k}-\frac{[k/2]}{[-k/2]}\frac{q^{-k}}{p^k}\right)-q^{N_{\Delta}}\left( q^{-k}-\frac{[k/2]}{[-k/2]}\frac{p^{-k}}{q^k}\right)  \right\rbrace ^{-1},
	\end{equation}
	and 
	\begin{equation}
	[k/2]=-(pq)^{k/2}[-k/2]
	\end{equation}
	giving
	\begin{equation}
	\chi_{ko}(p,q)= \left(p^{-k/2}+q^{-k/2} \right)^{-1} \left(p^{N_{\Delta}-k/2}- q^{N_{\Delta}-k/2}\right)^{-1}.
	\end{equation}
	Hence, 
	\begin{equation}\label{h1}
	\hat{X}_k =(p-q)\mathcal{R}\left(p^{-k},q^{-k}\right)\left(p^{-k/2}+q^{-k/2} \right)^{-1} \left(p^{N_{\Delta}-k/2}- q^{N_{\Delta}-k/2}\right)^{-1} .
	\end{equation}
	By analogy, we obtain
	\begin{equation}\label{h2}
	\hat{Y}_k =-(p-q)\mathcal{R}\left(p^{-k},q^{-k}\right)\left(p^{k/2}+q^{k/2} \right)^{-1} \left(p^{N_{\Delta}-k/2}- q^{N_{\Delta}-k/2}\right)^{-1} .
	\end{equation}
	Substituting  (\ref{h1}) and (\ref{h2}) in  (\ref{g1}) provides (\ref{g2}). 
	\item For $\Delta=2$, we have
		\begin{equation}
	\chi_{ko}(p,q)= \left\lbrace p^{N_{\Delta}}\left( p^{-k}-\frac{[2k]}{[k]}\frac{q^{-k}}{p^k}\right)-q^{N_{\Delta}}\left( q^{-k}-\frac{[2k]}{[k]}\frac{p^{-k}}{q^k}\right)  \right\rbrace ^{-1},
	\end{equation}
	and 
	\begin{equation}
	[2k]=(p^k + q^k)[k].
	\end{equation}
	Then,
	\begin{equation}
	\chi_{ko}(p,q)= -(pq)^k \left(p^{N_{\Delta}+k}- q^{N_{\Delta}+k}\right)^{-1}.
	\end{equation}
	The coefficient $\hat{X}_k$ and $\hat{Y}_k$ are given by:
	\begin{equation}
	\hat{X}_k =-(p-q)(pq)^k \left(p^{N_{\Delta}+k}- q^{N_{\Delta}+k}\right)^{-1}\mathcal{R}\left(p^{-k},q^{-k}\right) ,
	\end{equation}

	\begin{equation}
	\hat{Y}_k =-(p-q)(p^k + q^k) \left(p^{N_{\Delta}+k}- q^{N_{\Delta}+k}\right)^{-1}\mathcal{R}\left(p^{-k},q^{-k}\right),
	\end{equation}
	and the result follows. $\cqfd$
\end{enumerate}
{
\begin{proposition}
	 The $\mathcal{R}(p,q)$-deformed central charge $\tilde{C}^R_n(p,q)$
 is provided by:
	\begin{equation}\label{ce1}
	\tilde{C}^R_n(p,q) = C(p,q)(p^n + q^n)^{-1}(pq)^{\frac{N_{\Delta}}{2}+n}\mathcal{R}(p^{n-1},q^{n-1})\mathcal{R}(p^{n},q^{n})\mathcal{R}(p^{n+1},q^{n+1}),
	\end{equation}
	where
 $p,$ $q$  are two real numbers verifying  $0<q<p\leq 1,$ $C(p,q)$ is a function of $(p,q),$  and $\alpha^{\Delta}_{nmk}(p,q)$ is given by (\ref{J15}).
\end{proposition}}
{\it Proof:} 
According to the relations (\ref{J18}) and (\ref{J19}), we obtain

\begin{eqnarray}
[\tilde{L}_n, [\tilde{L}_m, \tilde{L}_k ]
&=& \mathcal{R}(p^{m-k},q^{m-k})\mathcal{R}(p^{-m-k{+}n},q^{-m-k{+}n})\tilde{L}_{n+m+k}\nonumber\\
&+& \mathcal{R}(p^{m-k},q^{m-k})\delta_{n+k+m,0}\tilde{\Gamma}(N_{\Delta})C^R_n(p,q),
\end{eqnarray}
and by analogy 
\begin{eqnarray}
[\tilde{L}_m, [\tilde{L}_k, \tilde{L}_n ]]&=& \mathcal{R}(p^{-n+k},q^{-n+k})\mathcal{R}(p^{-n-k+m},q^{-n-k+m})\tilde{L}_{n+m+k}\nonumber\\
&+& \mathcal{R}(p^{-n+k},q^{-n+k})\delta_{n+k+m,0}\tilde{\Gamma}(N_{\Delta})C^R_m(p,q),
\end{eqnarray}
and 
\begin{eqnarray}
[\tilde{L}_k, [\tilde{L}_n, \tilde{L}_m ]] &=& \mathcal{R}(p^{n-m},q^{n-m})\mathcal{R}(p^{-n-m+k},q^{-n-m+k})\tilde{L}_{n+m+k}\nonumber\\
&+& \mathcal{R}(p^{n-m},q^{n-m})\delta_{n+k+m,0}\tilde{\Gamma}(N_{\Delta})C^R_k(p,q).
\end{eqnarray}
Using the $\mathcal{R}(p,q)-$ deformed Jacobi identity{, we obtain
\begin{equation}\label{ce}
\displaystyle
\sum_{(u,v,l)\in\mathcal{C}(n,m,k)}\frac{(pq)^{-l}(p^u + q^u)}{\alpha^{\Delta}_{uvl}(p,q)}\mathcal{R}(p^{l-v},q^{l-v})\delta_{u+v+l,o}C^R_u(p,q) =0,
\end{equation}
where $\mathcal{C}(n,m,k)$ denotes the
cyclic permutation of $(n,m,k),$
leading to the following form of $C^R_n(p,q):$
\begin{equation}\label{ce2}
C^R_n(p,q) = C(p,q)(p^n + q^n)^{-1}(pq)^{n}\mathcal{R}(p^{n-1},q^{n-1})\mathcal{R}(p^{n},q^{n})\mathcal{R}(p^{n+1},q^{n+1}),
\end{equation}
with the solution of (\ref{g2})  given by:
\begin{equation}\label{g4}
\tilde{\Gamma}(N_{\Delta})=(pq)^{N_{\Delta}/2}.
\end{equation}
Then, using the relations (\ref{eg}), (\ref{ce2}),  and (\ref{g4}), we obtain the required result.
$\cqfd$}
\begin{theorem}
{Let $\tilde{L}_n$, $n\in\mathbb{Z},$ be the deformed generators and $\Delta$ a conformal dimension. Then, the $\mathcal{R}(p,q)$-deformed Virasoro algebra is driven by the following commutation relations }
	
	\begin{equation}\label{J21}
	[\tilde{L}_n, \tilde{L}_m]_{\hat{X},\hat{Y}}=\mathcal{R}(p^{n-m},q^{n-m})\tilde{L}_{n+m} + \delta_{n+m,0}\tilde{C}^R_n(p,q),
	\end{equation}
	where $\tilde{C}^R_n(p,q)$ given   by  (\ref{ce1}) is the central charge commuting with all $\tilde{L}_n,$ i.e.
	
	\begin{equation}\label{J22}
	[\tilde{L}_k, \tilde{C}^R_n(p,q)]_{\hat{X}_k,\hat{Y}_k}=0,
	\end{equation}
	with $\hat{X}_k$, $\hat{Y}_k $ 
 furnished   by (\ref{5}).
\end{theorem}
Note that the result obtained by Chakrabarti et {\it al} \cite{CJ} can be retrieved here by taking $\mathcal{R}(s,t)=(p-q)^{-1}(s^{-1}-t^{-1})$.

\begin{remark}
	The next particular cases are also worthy of attention. In each case, we give the commutation relation and the central element.
	{
	\begin{enumerate}	
		\item The $\mathcal{R}(p,q)$-deformed \textbf{Jagannathan-Srinivasa} \cite{JS}  Virasoro algebra:
		\begin{equation}
		[\tilde{L}_n, \tilde{L}_m]_{\hat{X},\hat{Y}}=[n-m]\tilde{L}_{n+m} + \delta_{n+m,0}\tilde{C}^R_n(p,q),
		\end{equation}
		 with 
		\begin{equation}\label {A2}
		\left \{
		\begin{array}{l}
		{\hat{X}}=(p^{n-m}-q^{n-m})\chi_{nm}(p,q) \\
		\\
		{\hat{Y}_{\Delta}}=(p^{n-m}-q^{n-m})\chi_{mn}(p,q).
		\end{array}
		\right. 
		\end{equation}
		\begin{equation}
		\tilde{C}^R_n(p,q) = C(p,q)(pq)^{\frac{N_{\Delta}}{2}+n}(p^n + q^n)^{-1}[n-1][n][n+1].
		\end{equation}
\item The $\mathcal{R}(p,q)$-deformed \textbf{Chakrabarti- Jagannathan}\cite{Chakrabarti&Jagan}  Virasoro algebra: 
\begin{equation}
[\tilde{L}_n, \tilde{L}_m]_{\hat{X},\hat{Y}}=[n-m]_{p^{-1},q}\tilde{L}_{n+m} + \delta_{n+m,0}\tilde{C}^R_n(p,q),
\end{equation}
where
\begin{equation}
\left \{
\begin{array}{l}
{\hat{X}}=(p-q)[n-m]_{p^{-1},q}\chi_{nm}(p,q) \\
\\
{\hat{Y}_{\Delta}}=(p-q)[n-m]_{p^{-1},q}\chi_{mn}(p,q)\\
\\
	\tilde{C}^R_n(p,q) = C(p,q)(pq)^{\frac{N_{\Delta}}{2}+n}(p^n + q^n)^{-1}[n-1]_{p^{-1},q}[n]_{p^{-1},q}[n+1]_{p^{-1},q}.
\end{array}
\right. 
\end{equation}
\item The $\mathcal{R}(p,q)$-deformed \textbf{Generalized $q$-Quesne} \cite{Hounkonnou&Ngompe07a} Virasoro algebra:
\begin{equation}
[\tilde{L}_n, \tilde{L}_m]_{\hat{X},\hat{Y}}=[n-m]^Q_{p,q}\tilde{L}_{n+m} + \delta_{n+m,0}\tilde{C}^R_n(p,q),
\end{equation}
where 
\begin{equation}
\left \{
\begin{array}{l}
{\hat{X}}=(p-q)[n-m]^Q_{p,q}\chi_{nm}(p,q) \\
\\
{\hat{Y}_{\Delta}}=(p-q)[n-m]^Q_{p,q}\chi_{mn}(p,q)\\
\\
\tilde{C}^R_n(p,q) = C(p,q)(pq)^{\frac{N_{\Delta}}{2}+n}(p^n + q^n)^{-1}[n-1]^Q_{p,q}[n]^Q_{p,q}[n+1]^Q_{p,q}.
\end{array}
\right. 
\end{equation}
\item The $\mathcal{R}(p,q)$-deformed \textbf{Hounkonnou-Ngompe generalized} \cite{Hounkonnou&Ngompe07a} Virasoro algebra:
\begin{equation}
[\tilde{L}_n, \tilde{L}_m]_{\hat{X},\hat{Y}}=[n-m]^{\mu,\nu}_{p, q, g}\tilde{L}_{n+m} + \delta_{n+m,0}\tilde{C}^R_n(p,q),
\end{equation}
with
\begin{equation}
\left \{
\begin{array}{l}
{\hat{X}}=(p-q)[n-m]^{\mu,\nu}_{p, q, g}\chi_{nm}(p,q) \\
\\
{\hat{Y}_{\Delta}}=(p-q)[n-m]^{\mu,\nu}_{p, q, g}\chi_{mn}(p,q)\\
\\
\tilde{C}^R_n(p,q) = C(p,q)(pq)^{\frac{N_{\Delta}}{2}+n}(p^n + q^n)^{-1}[n-1]^{\mu,\nu}_{p, q, g}[n]^{\mu,\nu}_{p, q, g}[n+1]^{\mu,\nu}_{p, q, g}.
\end{array}
\right. 
\end{equation}
\item The $\mathcal{R}(p,q)$-deformed \textbf{Hounkonnou-Bukweli} \cite{HB1} Virasoro algebra:
\begin{equation}
[\tilde{L}_n, \tilde{L}_m]_{\hat{X},\hat{Y}}=g(p,q)\left( \frac{q^{\nu}}{p^{\mu}}\right)^{n-m}[n-m]^Q_{p, q, g}\tilde{L}_{n+m} + \delta_{n+m,0}\tilde{C}^R_n(p,q),
\end{equation}
where 
\begin{equation}
\left \{
\begin{array}{l}
{\hat{X}}=g(p,q)(p-q)\left( \frac{q^{\nu}}{p^{\mu}}\right)^{n-m}[n-m]^Q_{p, q, g}\chi_{nm}(p,q) \\
\\
{\hat{Y}_{\Delta}}=g(p,q)(p-q)\left( \frac{q^{\nu}}{p^{\mu}}\right)^{n-m}[n-m]^Q_{p, q, g}\chi_{mn}(p,q)\\
\\
C^R_n(p,q) = C(p,q)g(p,q)^3\frac{(pq)^{\frac{N_{\Delta}}{2}+n}}{(p^n + q^n)}\left( \frac{q^{\nu}}{p^{\mu}}\right)^{3n}[n-1]^Q_{p, q, g}[n]^Q_{p, q, g}[n+1]^Q_{p, q, g}.
\end{array}
\right. 
\end{equation}
\end{enumerate}}
\end{remark}
\section{Deformation of conformal algebra  with $\Delta=1$}
In this section, we consider the algebra given by the relation (\ref{P1}) for $\Delta=1,$ and derive the relation between the $\mathcal{R}(p,q)$-deformed Virasoro algebra and the $\mathcal{R}(p,q)$- KdV equation. Relevant particular cases of the  $(p,q)-$deformed KdV equation are also deduced.
\begin{proposition}
	Let $\phi(z)$ be the fields of the conformal dimension $\Delta=1,$ and $p$ and $q$ two real numbers verifying $0<q<p\leq 1$. Then the $\mathcal{R}(p,q)-$ deformed generators
$\mathcal{L}^1_n $ such that
\begin{equation}\label{d2}
\mathcal{L}^1_n \phi(z)=[z\partial_z +1]z^nw^1_n(p,q)\phi(z)
\end{equation}
 define an algebra 
obeying the following commutation relation
	\begin{equation}\label{d4}
	[\mathcal{L}^{1}_n , \mathcal{L}^{1}_m]_{\hat{x},\hat{y}}\phi(z)=\mathcal{R}(p^{n-m},q^{n-m})q^{N_{1}-m}\mathcal{L}^{1}_{n+m}\phi(z),
	\end{equation}
	with 
	\begin{equation}\label {d5}
	\left \{
	\begin{array}{l}
	\displaystyle
	\hat{x}=\hat{\chi}_{nm}(p,q)\\
	\\
	
\hat{y}=p^{m-n}\hat{\chi}_{nm}(p,q),
	\end{array}
	\right. 
	\end{equation}
	and
	
	\begin{equation}\label{d6}
	\hat{\chi}_{nm}(p,q)=\frac{\mathcal{R}(p^{n-m},q^{n-m})}{[m-n]}\frac{w^{1}_{n+m}(p,q)}{w^{1}_n(p,q)w^{1}_m(p,q)}.
	\end{equation}
\end{proposition}
{\it Proof:} It follows from the  definition of the commutator:
\begin{equation}\label{d7}
[\mathcal{L}^{1}_n , \mathcal{L}^{1}_m]_{\hat{x},\hat{y}}\phi(z)=\hat{x}\mathcal{L}^{1}_n\mathcal{L}^{1}_m\phi(z) - \hat{y}\mathcal{L}^{1}_m\mathcal{L}^{1}_n\phi(z).
\end{equation}
$ \cqfd$ 
 
 Let us now rewrite the $\mathcal{R}(p,q)$-deformed generators as follows:
 
 \begin{equation}\label{d13}
 \tilde{L}^1_n\phi(z)=q^{-N_1}\mathcal{L}^{1}_n\phi(z),
 \end{equation}
 or, equivalently, by using Proposition \ref{p1}:
 
 \begin{equation}\label{d14}
  \tilde{L}^1_n\phi(z)=(p-q)^{-1}\left((\frac{p}{q}) \phi(p/qz)-\phi(z)\right)z^nw^1_n(p,q).
 \end{equation}
 
 \begin{lemma}\label{La}
 	The $\mathcal{R}(p,q)-$generators $\tilde{L}^1_n$ defined by (\ref{d13}) satisfy the following relations:
 \begin{equation}\label{d15}
 [\tilde{L}^1_n , \tilde{L}^1_m]_{x,y}\phi(z)=\mathcal{R}(p^{n-m},q^{n-m})\tilde{L}^1_{n+m}\phi(z)
 \end{equation}
 where
  
\begin{equation}\label {d16}
\left \{
\begin{array}{l}
\displaystyle
x=q^{-n+m}\hat{\chi}_{nm}(p,q)\\
\\

y=p^{-n+m}\hat{\chi}_{mn}(p,q)
\end{array}
\right. 
\end{equation}
or, equivalently,
\begin{equation}\label{d17}
[\tilde{L}^1_n , \tilde{L}^1_m]\phi(z)=[m-n]p^{N_1-m}q^{-N_1+n}K^1_{nm}(p,q)\tilde{L}^1_{n+m}\phi(z).
\end{equation}
with
\begin{equation}\label{d9}
K^1_{nm}(p,q)= \frac{w^1_n(p,q)w^1_m(p,q)}{w^1_{n+m}(p,q)}.
\end{equation}
\end{lemma}
\begin{remark} Note that:
	\begin{enumerate}
\item Using the symmetry $p\longrightarrow q$ and $q\longrightarrow p^{-1}$, we obtain  the result given in \cite{CJ} from the relations (\ref{d16}) and (\ref{d17}).
\item  The $\mathcal{R}(p,q)-$ deformed $su(1,1)$ subalgebra is generated by the  commutation relations:
\begin{itemize}
\item \begin{equation}\label{d23}
[\tilde{L}^1_0 , \tilde{L}^1_1]_{x,y}\phi(z)=\mathcal{R}(p^{-1},q^{-1})\tilde{L}^1_{1}\phi(z)
\end{equation}
where 

\begin{equation}\label{d24}
x=q\hat{\chi_{01}(p,q)}\quad\mbox{and}\quad y=p\hat{\chi}_{01}(p,q)
\end{equation}

\item \begin{equation}\label{d25}
[\tilde{L}^1_{-1} , \tilde{L}^1_0]_{x,y}\phi(z)=\mathcal{R}(p^{-1},q^{-1})\tilde{L}^1_{-1}\phi(z)
\end{equation}
where 

\begin{equation}\label{d26}
x=q\hat{\chi}_{-10}(p,q)\quad\mbox{and}\quad y=p\hat{\chi}_{-10}(p,q)
\end{equation}
\item \begin{equation}\label{d27}
[\tilde{L}^1_{-1} , \tilde{L}^1_1]\phi(z)= [2]p^{N_1-1}q^{-N_1-1}K^1_{-10}(p,q)\tilde{L}^1_{0}\phi(z).
\end{equation}
\end{itemize}
\item The commutation relation  (\ref{d27}) defines an $\mathcal{R}(p,q)$-deformation of the   Witten algebra \cite{W} in the case of the vertex models.
\item  The $\mathcal{R}(p,q)$ generators defined by 
\begin{equation}\label{d28}
\tilde{\mathcal{L}}_n\phi(z): = \lambda^{-1}\tilde{L}^1_n\phi(z).
\end{equation} 
 satisfy the commutation relation
	
	\begin{equation}\label{d29}
	[\tilde{\mathcal{L}}_n , \tilde{\mathcal{L}}_m]_{x,y}\phi(z) = \mathcal{R}(\theta^{m-n} , \theta^{n-m})\tilde{\mathcal{L}}_{n+m}\phi(z)
	\end{equation} 
	where
	
		\begin{equation}\label {d30}
		\left \{
		\begin{array}{l}
		\displaystyle
		x=\lambda^{m-n}\theta^{n-m}\tilde{\chi}_{nm}(\theta,\theta^{-1})\\
		\\
		
		y=\lambda^{m-n}\theta^{m-n}\tilde{\chi}_{nm}(\theta,\theta^{-1}).
		\end{array}
		\right. 
		\end{equation}
\end{enumerate}
\end{remark}
\subsection{$\mathcal{R}(p,q)$-deformed nonlinear equation}

Chaichian et {\it al} \cite{CPP} used the well known connection between the Virasoro algebra and the KdV equation to derive a $q-$ deformed KdV equation corresponding to a $q-$Virasoro algebra. Chakrabarti et {\it al}\cite{CJ} used a similar formalism to study the correlation between a $(p,q)-$Virasoro algebra and a $(p,q)$-KdV equation. The method used in these works
is based on the construction of a current defining a bi-Hamiltonian structure which satisfies a nonlinear evolution equation. We follow the same  procedure to derive a $\mathcal{R}(p,q)$-deformed nonlinear differential equation 
corresponding to the algebra (\ref{d34}). 

The central extension of the algebra (\ref{d15}) is generated by the commutation relation

\begin{equation}\label{d31}
[\tilde{L}^1_n , \tilde{L}^1_m]_{x,y}=\mathcal{R}(p^{n-m},q^{n-m})\tilde{L}^1_{n+m} + \delta_{n+m,o}\tilde{C}^{R}_n(p,q),
\end{equation}
where $x$ and $y$ are given by (\ref{d16}),  and, for all $\tilde{L}^1_k,$
\begin{equation}\label{d32}
[\tilde{L}^1_k,\tilde{C}^{R}_n(p,q)]_{x_k,y_k}=0,
\end{equation}
with $x_k=q^{-k}\chi_{ko}$ and $y_k=p^{-k}\chi_{ok}$.

We consider now the generators defined as follows:
\begin{equation}\label{d33}
t_n := \left(\frac{q}{p}\right)^{N_1}\tilde{L}^1_n.
\end{equation}
Then, using the algebra (\ref{d17}), the generators (\ref{d33}) satisfy the following deformed Virasoro algebra

\begin{equation}\label{d34}
[t_n , t_m]=[m-n]p^{N_1-m}q^{-N_1+n}K^1_{nm}(p,q)t_{n+m}+ \delta_{n+m,o}\tilde{C}^R_n(p,q),
\end{equation}
where the deformed central charge commuting with all generators $t_i,$ $i\in\mathbb{Z},$ is expressed as:
\begin{equation}
\tilde{C}^R_n(p,q)= C(p,q)p^{\frac{N_1}{2}+n}q^{\frac{N_1}{2}-n}(p^n + q^n)^{-1}\mathcal{R}(p^{n-1},q^{n-1})\mathcal{R}(p^{n},q^{n})\mathcal{R}(p^{n+1},q^{n+1}).
\end{equation}
\begin{definition}
	Let $t_n,$ $n\in\mathbb{Z},$ be the generators considered in (\ref{d33}), and $x\in\mathbb{R}.$  Then the $\mathcal{R}(p,q){-}${deformed} current is given by:
	\begin{equation}
	v(x):=\sum_{n\in\mathbb{Z}}t_ne^{-inx}.
	\end{equation}
\end{definition}
This definition is consistent with 
{ the $(p,q)-$deformed current  given in \cite{CJ}:
	\begin{equation}\label{pc}
		u(x):=\sum_{n\in\mathbb{Z}}e_ne^{-inx}
	\end{equation}
	where $e_n$ are  $(p,q)-$deformed generators.}
Using the relation between the $\mathcal{R}(p,q)-$  and $(p,q)-$ generators, we can re-express  the $\mathcal{R}(p,q)-$ deformed current as follows:
{
\begin{equation}
v(x):=\sum_{n\in\mathbb{Z}}w^1_n(p,q)e_ne^{-inx}.
\end{equation}
After computation, we obtain
\begin{equation}
v(x)= \alpha\,u(x)- \gamma_{nm}(x)
\end{equation} where  $\displaystyle\alpha=\sum_{n\in\mathbb{Z}}w^1_n(p,q)$, $\displaystyle\gamma_{nm}(x)=\sum_{n\neq m}w^1_n(p,q)e_m\,e^{-imx}$ and $u(x)$ is given by (\ref{pc}). } 
{
Then, we arrive at the following commutation relation 
\begin{eqnarray}
	\Big[v(x),v(y)\Big]&=&\alpha^2\,\Big[u(x),u(y)\Big]. 
\end{eqnarray}
The latter yields, after some algebra,
\begin{eqnarray}
\frac{1}{2\pi\,i}\Big[v(x),v(y)\Big]
&=& \alpha^2\,\frac{\theta}{2\sin \epsilon}\Big( e^{-2\epsilon\partial_x}v(x)-v(x)e^{2\epsilon\partial_x}\Big)\lambda^{-2N_1}\delta(x-y)
\nonumber\\
&-&\theta^{3}\frac{\sinh \epsilon\partial_x }{\sinh 2\epsilon\partial_x }\frac{\sinh \epsilon(\partial_x+i)\sinh \epsilon\partial_x \sinh \epsilon(\partial_x -i) }{\sin ^3\epsilon}\nonumber\\
&\times& \lambda^{-2N_1}\delta(x-y),
\end{eqnarray}
where $\lambda$ and  $\theta=e^{-i\epsilon},$ $\epsilon\in\mathbb{R}^{*}$  are given by the relation (\ref{e9}).
This  generates the $\mathcal{R}(p,q)-$deformed KdV equation  as follows:
\begin{eqnarray}
\frac{dv}{dx}&=&\frac{\alpha^2\theta}{4\sin \epsilon}\Big( e^{-2\epsilon\partial_x}v(x)-v(x)e^{2\epsilon\partial_x}\Big)\Big(\lambda^{-2N_1}v(x)+v(x)\lambda^{-2N_1}\Big)\nonumber\\
&-& \alpha^2\frac{\theta^{3}}{2}\frac{\sinh \epsilon\partial_x }{\sinh 2\epsilon\partial_x }\frac{\sinh \epsilon(\partial_x+i)\sinh \epsilon\partial_x \sinh \epsilon(\partial_x -i) }{\sin ^3\epsilon}\nonumber\\
&\times& \Big(\lambda^{-2N_1}v(x)+v(x)\lambda^{-2N_1}\Big),
\end{eqnarray} }
which
 reduces to the Chakrabarti et {\it al}\cite{CJ} KdV equation in the particular case of  $\mathcal{R}(s,t)=(p-q)^{-1}(s^{-1}-t^{-1})$.

\begin{remark} We can easily deduce the nonlinear differential  equations associated to particular algebras described in the previous sections.
Without loss of generality, { for the computation of $\alpha,$ we consider only the  $\mathbb{Z}_{+}$ part, and find the following results:
\begin{enumerate} 
		\item The \textbf{Jagannathan-Srinivasa} \cite{JS} deformed KdV equation:
	\begin{eqnarray}
	\frac{dv}{dx}&=&\frac{\alpha^2\theta}{4\sin \epsilon}\Big( e^{-2\epsilon\partial_x}v(x)-v(x)e^{2\epsilon\partial_x}\Big)\Big(\lambda^{-2N_1}v(x)+v(x)\lambda^{-2N_1}\Big)\nonumber\\
	&-&\alpha^2 \frac{\theta^{3}}{2}\frac{\sinh \epsilon\partial_x }{\sinh 2\epsilon\partial_x }\frac{\sinh \epsilon(\partial_x+i)\sinh \epsilon\partial_x \sinh \epsilon(\partial_x -i) }{\sin ^3\epsilon}\nonumber\\
	&\times& \Big(\lambda^{-2N_1}v(x)+v(x)\lambda^{-2N_1}\Big),
	\end{eqnarray}
	where $\alpha = \frac{pq}{pq-1},$ with $|pq|<1.$
		\item  The deformed \textbf{Chakrabarty et {\it al}} \cite{Chakrabarti&Jagan} 
	KdV equation:
	\begin{eqnarray}
	\frac{dv}{dx}&=&\frac{\alpha^2\theta}{4\sin \epsilon}\Big( e^{-2\epsilon\partial_x}v(x)-v(x)e^{2\epsilon\partial_x}\Big)\Big(\lambda^{-2N_1}v(x)+v(x)\lambda^{-2N_1}\Big)\nonumber\\
	&-& \alpha^2\frac{\theta^{3}}{2}\frac{\sinh \epsilon\partial_x }{\sinh 2\epsilon\partial_x }\frac{\sinh \epsilon(\partial_x+i)\sinh \epsilon\partial_x \sinh \epsilon(\partial_x -i) }{\sin ^3\epsilon}\nonumber\\
	&\times& \Big(\lambda^{-2N_1}v(x)+v(x)\lambda^{-2N_1}\Big),
	\end{eqnarray}
	where
	\begin{equation}
      \alpha=\frac{p-q}{p^{-1}-q}\big(\frac{q}{q-p} - \frac{pq}{1-pq}- \frac{p^2}{1 -p^2}\big)
	\end{equation}
	 $|\frac{p}{q}|< 1$ and $|pq|<1.$
		\item  The \textbf{generalized Quesne} \cite{Hounkonnou&Ngompe07a} KdV equation is given by :	
	\begin{eqnarray}
	\frac{dv}{dx}&=&\frac{\alpha^2\theta}{4\sin \epsilon}\Big( e^{-2\epsilon\partial_x}v(x)-v(x)e^{2\epsilon\partial_x}\Big)\Big(\lambda^{-2N_1}v(x)+v(x)\lambda^{-2N_1}\Big)\nonumber\\
	&-& \alpha^2\frac{\theta^{3}}{2}\frac{\sinh \epsilon\partial_x }{\sinh 2\epsilon\partial_x }\frac{\sinh \epsilon(\partial_x+i)\sinh \epsilon\partial_x \sinh \epsilon(\partial_x -i) }{\sin ^3\epsilon}\nonumber\\
	&\times& \Big(\lambda^{-2N_1}v(x)+v(x)\lambda^{-2N_1}\Big),
	\end{eqnarray}
	with 
	\begin{equation}
	\alpha =\frac{p-q}{q-p^{-1}}\big(-\frac{q^2 }{1-q^2} + \frac{p}{p-q}- \frac{pq}{1-pq}\big)
	\end{equation}
	 $|\frac{q}{p}|< 1$ and $|pq|<1.$
		\item  The  deformed \textbf{Hounkonnou-Ngompe generalized}\cite{Hounkonnou&Ngompe07a} KdV equation:
		\begin{eqnarray}
		\frac{dv}{dx}&=&\frac{\alpha^2\theta}{4\sin \epsilon}\Big( e^{-2\epsilon\partial_x}v(x)-v(x)e^{2\epsilon\partial_x}\Big)\Big(\lambda^{-2N_1}v(x)+v(x)\lambda^{-2N_1}\Big)\nonumber\\
		&-& \alpha^2\frac{\theta^{3}}{2}\frac{\sinh \epsilon\partial_x }{\sinh 2\epsilon\partial_x }\frac{\sinh \epsilon(\partial_x+i)\sinh \epsilon\partial_x \sinh \epsilon(\partial_x -i) }{\sin ^3\epsilon}\nonumber\\
		&\times& \Big(\lambda^{-2N_1}v(x)+v(x)\lambda^{-2N_1}\Big),
		\end{eqnarray}
		where
			\begin{equation}
		\alpha =g(p,q)\frac{p-q}{q-p^{-1}}\Big(\frac{q^{\nu}}{p^{\mu}-q^{\nu}}+ \frac{q^{\nu+1}}{p^{\mu+1}-q^{\nu+1}}-\frac{q^{\nu+2}}{p^{\mu-1}-q^{\nu+2}}+\frac{q^{\nu+2}}{p^{\mu}-q^{\nu+2}}\Big),
		\end{equation}
		with $|\frac{q}{p}|<1$
	\end{enumerate}}
\end{remark}

\section{$\mathcal{R}(p,q)-$ deformed energy-momentum tensor for the case
 $\Delta=2$}
In this section, we  compute  the $\mathcal{R}(p,q)-$deformed energy-momentum tensor for the conformal dimension  $\Delta=2$.
\subsection{$(p,q)-$ deformed energy-momentum tensor}
Let us start with the simplest case of the  $(p,q)$-deformed algebra given in \cite{CJ} for  $\Delta =2$. The corresponding generators are given as follows:

\begin{equation}\label{t1}
L^2_n \phi_2(z) = z^{-(n+1)}D_{p,q}(z^{2(n+1)}\phi_2(z)),
\end{equation}
 satisfying the commutation relation

\begin{eqnarray}\label{t2}
[L^2_n, L^2_m]_{x_2,y_2}\phi(z)&=& K\left\lbrace p^{N_2}(x_2p^{-n}-y_2p^{-m})-q^{N_2}(x_2q^{-n}-y_2q^{-m}) \right\rbrace  L^2_{n+m}\phi(z)\nonumber\\
&=& (pq)^n[m-n](p^{N_2}+q^{N_2})L^2_{n+m}\phi(z),
\end{eqnarray}
with 

\begin{equation}\label {t3}
\left \{
\begin{array}{l}
\displaystyle
y_2=(pq)^n(p^m + q^m)\\
\\

y_2=(pq)^m(p^n + q^n)\\
\\
K=(p-q)^{-1}.
\end{array}
\right. 
\end{equation}
According to \cite{BPZ} and \cite{Z}, the energy-momentum $\mathcal{T}(z)$ has the conformal dimension two but does not transform as a primary field \cite{CILPP}. In the undeformed case, 

\begin{equation}\label{t4}
\mathcal{T}(z)\longrightarrow ( \partial_z\phi(z))^2 +\phi(z) + c\left(\frac{\partial^{(3)}_z\phi(z)}{\partial_z}-3/2 ( \frac{\partial^{(2)}_z\phi(z)}{\partial_z\phi(z)})^2\right),
\end{equation} 
where $c$ is the central charge and $\partial^{(n)}(z)$ is the $n-$th order derivative. The infinitesimal form is given by: 

\begin{equation}\label{t5}
\delta_{\epsilon}\mathcal{T}(z) = (\epsilon(z)\partial_z + 2\partial\epsilon(z))\mathcal{T}(z) + c\partial_z^{(3)}\epsilon(z).
\end{equation}
Putting $\epsilon(z)=z^{n+1}$, we have 

\begin{equation}\label{t6}
\delta_n\mathcal{T}(z)=l_n\mathcal{T}(z) + cg(n)z^{n-2},
\end{equation}
where 

\begin{equation}\label{t7}
l_n\phi(z)=[z\partial_z + n+2]z^n\phi(z)
\end{equation}
and $g(n)=(n-1)n(n+1)$.

If the equation (\ref{t7}) satisfies

\begin{equation}\label{t8}
[\delta_m, \delta_n]\mathcal{T}(z)=(m-n)\delta_{n+m}\mathcal{T}(z), 
\end{equation}
then the central term can be obtained.
 
The $(p,q)-$ analogue of equation (\ref{t6}) is written as:

\begin{equation}\label{t9}
\delta^{p,q}_n\mathcal{T}_{p,q}(z)=L^2_n\mathcal{T}_{p,q}(z) + c_n(p,q)z^{n-2},
\end{equation}

where $\mathcal{T}_{p,q}(z)$ is the $(p,q)-$ deformed energy-momentum tensor and $L^2_n$ is defined by (\ref{t1}). We consider now $c_n(p,q)$ as the central extension term for the $(p,q)-$ deformed conformal algebra. Using (\ref{t2}), we obtain 

\begin{equation}\label{t10}
[\delta^{p,q}_m , \delta^{p,q}_n]_{x_2,y_2}\mathcal{T}_{p,q}(z)=(pq)^n[m-n](p^{N_2} + q^{N_2})\delta^{p,q}_{n+m}\mathcal{T}_{p,q}(z),
\end{equation}
where $x_2$ and $y_2$ are given by equation (\ref{t3}).

According  to the relations (\ref{t9}), and (\ref{t10}),  and using (\ref{t2}), we get
\begin{equation}\label{t11}
\nu_{nm}(p,q) c_m(p,q)- \mu_{nm}(p,q) c_n(p,q) = \alpha_{nm}(p,q) c_{n+m}(p,q),
\end{equation}
with $\nu_{nm}(p,q)=(pq)^n(p^m + q^m)[2n+m]$, $\mu_{nm}(p,q)=(pq)^{m-n}(p^n + q^n)[2m+n]$ and $ \alpha_{nm}(p,q)
= [m-n](p^{m+n}+q^{m+n})$.
Putting $m=1$ and $n=0$ in the equation (\ref{t11}), we have $c_0(p,q)=0$. But if $m=-n$, we get $c_n(p,q) = c_{-n}(p,q)$. Thus, we have $c_1 (p,q)\neq 0$.

Now  defining the tensor such that $c_1(p,q)=0$ imposes to shift
 $\mathcal{T}_{p,q}(z)$ as follows:

\begin{equation}\label{t12}
\hat{\mathcal{T}}_{p,q}(z) = \mathcal{T}_{p,q}(z) + \frac{\beta(p,q)}{[2]}z^{-2},
\end{equation}
where $\beta$ is a constant depending on the parameters $p$ and $q$. Thus the function $c_n(p,q)$ takes the following form

\begin{equation}\label{t13}
\hat{c}_n(p,q)= c_n(p,q) +  \beta(p,q)\frac{[2n]}{[2]}.
\end{equation}
Hence , if we choose $\beta=-c_1(p,q)$, then $\hat{c}_1(p,q)=0$.
Finally, we get from (\ref{t11}) the following equation

\begin{equation}
(p^m + q^m)[m-2]\hat{c}_m(p,q) = (pq)^{-2}(p^{m-1}+q^{m-1})[m+1]\hat{c}_{m-1}(p,q.).
\end{equation}
Its solution is the $(p,q)-$deformed central extension given in \cite{CJ}.
\subsection{$\mathcal{R}(p,q)-$ deformed energy-momentum tensor}
The corresponding $\mathcal{R}(p,q)$ generators for $\Delta=2$ are given as follows:

\begin{equation}\label{t18}
\mathcal{L}^{(2)}_n\phi(z)=z^{-(1+n)}D_{\mathcal{R}(p,q)}(z^{2(1+n)}\phi(z)),
\end{equation}
and satisfy the following relation
\begin{eqnarray}\label{t19}
[\mathcal{L}^{2}_n , \mathcal{L}^{2}_m]_{\tilde{X}_{2},\tilde{Y}_{2}}\phi(z)
&=& K\left\lbrace p^{N_{2}}(\tilde{X}_{2}p^{-n}-\tilde{Y}_{2}p^{-m}) - q^{N_{2}}(\tilde{X}_{2}q^{-n}-\tilde{Y}_{2}q^{-m}) \right\rbrace\nonumber\\
&\times& \mathcal{L}^{2}_{n+m}\phi(z)\nonumber\\
&=& (pq)^nK_{nm}(p,q)[m-n](p^{N_2}+q^{N_2})L^2_{n+m}\phi(z),
\end{eqnarray} 

where 
\begin{equation}\label {t20}
\left \{
\begin{array}{l}
\tilde{X}_{2}= (pq)^{n}(p^m + q^m)K_{nm}(p,q)\\
\\
\tilde{Y}_{2}=(pq)^{m}(p^n + q^n)K_{nm}(p,q)\\
\\
N_{2} =z\partial_z + 2\\
\\
K_{nm}(p,q)=\frac{w^{2}_{n+m}(p,q)}{w^2_n(p,q)w^2_m(p,q)} \\
\\
K=(p-q)^{-1}.
\end{array}
\right. 
\end{equation} 
According to the equation (\ref{t6}),
	the $\mathcal{R}(p,q)-$ deformed infinitesimal form can be given by:
	\begin{equation}\label{t21}
	\delta_n \mathcal{T}_{\mathcal{R}(p,q)}(z)=\mathcal{L}^{2}_n\mathcal{T}_{\mathcal{R}(p,q)}(z) + \tilde{C}^R_n(p,q)z^{n-2},
	\end{equation} 
	where  $\mathcal{L}^{2}_n$ are the deformed generators given by (\ref{t19}),  $\tilde{C}^R_n(p,q)$ is the central charge of the $\mathcal{R}(p,q)-$ deformed conformal algebra, and  $\mathcal{T}_{\mathcal{R}(p,q)}(z)$ is the $\mathcal{R}(p,q)-$ deformed energy-momentum tensor.
Therefore, 
	the $\mathcal{R}(p,q)-$ deformed infinitesimal form satisfies the following commutation relation
	\begin{equation}\label{t22}
	[\delta_m , \delta_n]_{\tilde{X}_{2},\tilde{Y}_{2}} \mathcal{T}_{\mathcal{R}(p,q)}(z) =(pq)^nK_{nm}(p,q)[m-n](p^{N_2}+q^{N_2})\delta_{n+m}\mathcal{T}_{\mathcal{R}(p,q)}(z), 
	\end{equation} 
	where $\tilde{X}_{2}$, $\tilde{Y}_{2}, $ and $K_{nm}(p,q)$ are given by (\ref{t20}),
which can be deduced by a straightforward computation. Furthermore,
using  the relations (\ref{t19}), (\ref{t21}),  and (\ref{t22}), we obtain 
\begin{equation}\label{t23}
\nu^R_{nm}(p,q) C^R_m(p,q)- \mu^R_{nm}(p,q) C^R_n(p,q) = \alpha^R_{nm}(p,q) C^R_{n+m}(p,q),
\end{equation}
where
 $\nu^R_{nm}(p,q)=(pq)^n(p^m + q^m)[2n+m]w^2_n(p,q),$\quad $\mu^R_{nm}(p,q)=(pq)^m(p^n + q^n)[2m+n]w^2_m(p,q),$ \quad and \quad $ \alpha^R_{nm}(p,q)
=(pq)^n [m-n](p^{m+n}+q^{m+n})$.

The $\mathcal{R}(p,q)-$deformed tensor is then  given by:
	\begin{equation}
	\tilde{\mathcal{T}}_{\mathcal{R}(p,q)}(z)= \mathcal{T}_{\mathcal{R}(p,q)}(z) + \frac{\gamma(p,q)}{[2]}z^{-2}.
	\end{equation}
\section{Concluding  remarks}
In this paper, we have constructed an $\mathcal{R}(p,q)-$ deformed conformal Virasoro algebra with an arbitrary conformal dimension $\Delta.$ 
Wellknown  deformed algebras, 
 investigated in the literature,
 have been deduced as particular cases. 
A special attention has been paid to the specific case of the conformal dimension $\Delta=1$ for its interesting properties. 
The $\mathcal{R}(p,q)-$ nonlinear differential  equation has been derived, and its link to  the $\mathcal{R}(p,q)-$ deformed Virasoro algebra has been  established. The    $\mathcal{R}(p,q)-$ deformed energy-momentum tensor,
 consistent with the obtained deformed central extension term,
 has been computed. 
 \section*{Acknowledgements}
This work is supported by TWAS Research Grant RGA No. 17 - 542 RG / MATHS / AF / AC \_G  -FR3240300147. The ICMPA-UNESCO Chair is in partnership with Daniel Iagolnitzer Foundation (DIF), France, and the Association pour la Promotion Scientifique de l'Afrique (APSA), supporting the development of mathematical physics in Africa.

\end{document}